\documentstyle[12pt,epsfig,colordvi]{article}
\let\section=\subsection     \let\subsection=\subsubsection                %%
\def\tr{{\rm tr} \,}

\begin{document}
\begin{center}
   {\large \bf Chiral symmetry and resonances in QCD}\\[2mm]
      M.F.M.~LUTZ$^{a,b}$ and E.E.~KOLOMEITSEV$^c$\\[5mm]
   {\small \it $^a$Gesellschaft f\"ur Schwerionenforschung (GSI) \\
   Postfach 110552, D-64220 Darmstadt, Germany  }
   \\[5mm]
   {\small \it $^b$ Institut f\"ur Kernphysik, TU Darmstadt\\
D-64289 Darmstadt, Germany}\\[5mm]
   {\small \it $^c$ The Niels Bohr Institute\\ Blegdamsvej
17, DK-2100 Copenhagen, Denmark}
\end{center}

\begin{abstract}\noindent
We study the formation of resonances in terms of the chiral SU(3)
Lagrangian. At leading order parameter-free predictions are
obtained for the scattering of Goldstone bosons off any hadron
once we insist on approximate crossing symmetry of the unitarized
scattering amplitude. A wealth of empirically established
resonances are recovered and so far unseen resonances are
predicted.  Here we review in some detail the properties of light
axial-vector mesons.
\end{abstract}

\section{Introduction}

A profound understanding of the nature of meson and baryon
resonances is one of the outstanding issues of QCD. The discovery
of meson and baryon resonance states with exotic quantum numbers
turns this  into a burning question deserving prime attention. The
goal is to establish a systematic approach which is capable of
describing and predicting the resonance spectrum of QCD
quantitatively. In this context the authors' radical conjecture
that meson and baryon resonances that do not belong to the
large-$N_c$ ground state of QCD \cite{LK02,LH02,LK03} should be
generated in terms of coupled-channel dynamics strongly questions
the quantitative applicability of the constituent-quark model to
the resonance spectrum of QCD. In more explicit terms we expect
that the resonance spectrum in the $(u,d,s)$-sector should be
described in terms of the baryon-octet $\frac{1}{2}^+$ and
baryon-decuplet $\frac{3}{2}^+$ fields together with the Goldstone
boson $0^-$ and vector meson $1^-$ fields. In the large-$N_c$ and
heavy-quark mass limit of QCD the latter baryon fields and also
the latter meson fields have degenerate masses. If this change of
paradigm is justified it requires to provide a systematic
explanation of the many meson and baryon resonances established
experimentally.

It is long known that the spectrum of light scalar mesons can be
successfully reproduced in terms of coupled-channel dynamics as
first pointed out by T\"ornqvist~\cite{Torn} and Van Beveren
\cite{Rupp86} in the 80s. In recent years this idea has been
systematized by applying the chiral Lagrangian (see e.g.
\cite{NVA02,NP02}). Guided by the argument that in the heavy-quark
mass limit one expects a degeneracy of the $0^+$ and $1^+$ meson
spectra it is natural to expect a similar description of the
axial-vector meson spectrum \cite{LK03}, to be reviewed in this
talk. Even earlier in the 60s a series of works
\cite{Wyld,Dalitz,Ball,Rajasekaran,Wyld2} predicted a wealth of
s-wave baryon resonance generated by coupled-channel dynamics.
Those works were based on a SU(3)-symmetric interaction Lagrangian
that is closely related to the leading order chiral Lagrangian.
For recent firm results on the spectrum of s- but also for the
first time on d-wave baryon resonances based on the chiral
Lagrangian see \cite{Granada,Copenhagen}. In the latter works it
was demonstrated that chiral SU(3) symmetry generates
parameter-free the $J^P\!=\!\frac{1}{2}^-$ and
$J^P\!=\!\frac{3}{2}^-$ baryon resonance spectrum. These results
are important because they demonstrate that chiral coupled-channel
theory is able to predict in which channels QCD forms resonance
states.

The spectrum of axial-vector mesons is obtained by studying the
scattering of Goldstone bosons off vector mesons using the leading
order chiral Lagrangian. Our results are parameter-free once we
insist on approximate crossing symmetry for unitarized scattering
amplitudes. We find that chiral symmetry predicts the existence of
the axial-vector meson resonances ($h_1(1170), h_1(1380)$,
$f_1(1285), a_1(1260)$, $b_1(1235)$, $K_1(1270)$, $K_1(1400)$). A
result analogous to the baryon sector
\cite{LK02,Granada,Copenhagen} is found: in the heavy SU(3) limit
with $m_{\pi, K, \eta} \simeq 500$ MeV and $m_{\rho, \omega, K^*,
\phi}\simeq 900$ MeV the resonance states turn into bound states
forming two degenerate octets and one singlet of the SU(3) flavor
group with masses  1360 MeV and 1280 MeV respectively. Taking the
light SU(3) limit with $m_{\pi, K, \eta} \simeq 140$ MeV and
$m_{\rho, \omega, K^*, \phi}\simeq 700$ MeV we do not observe any
bound-state nor resonance signals anymore. Since the leading order
interaction kernel scales with $N_c^{-1}$ the resonances disappear
also in the large-$N_c$ limit. Using physical masses a pattern
arises that compares surprisingly well with the empirical
properties of the $J^P\!=\!1^-$ meson resonances.

\section{The $\chi$-BS(3) approach}

The starting point to study the scattering of Goldstone bosons off
vector mesons is the chiral SU(3) Lagrangian. The leading order
Weinberg-Tomozawa interaction Lagrangian density \cite{Wein-Tomo}
\begin{eqnarray}
{\mathcal L}_{WT}(x) &=& -\frac{1}{16\,f^2}\,\tr \Big(
[\Phi^\mu(x)\,, (\partial^\nu \Phi_\mu(x) )]_- \,[\phi (x) ,
(\partial_\nu\,\phi(x))]_- \Big) \,, \label{WT-term}
\end{eqnarray}
describes the s-wave interaction of the Goldstone bosons field
$\phi$ with a massive vector-meson field $\Phi_\mu$. The parameter
$f\simeq 90$ MeV in (\ref{WT-term}) is known from the weak decay
process of the pions. In (\ref{WT-term}) we omit additional terms
that do not contribute to the on-shell scattering amplitude of
Goldstone bosons off vector-meson at tree-level.

As was emphasized in \cite{LK02} the chiral SU(3) Lagrangian
should not be used in perturbation theory except at energies
sufficiently below all thresholds. Though the infinite set of
irreducible diagrams can be successfully approximated by the
standard perturbative chiral expansion that is no longer true for
the infinite set of reducible diagrams once the energies are
sufficiently large to support hadronic scattering processes. From
an effective field theory point of view it is mandatory to sum the
reducible diagrams. This is naturally achieved by considering the
Bethe-Salpeter scattering equation. Since the intermediate vector
mesons have in part a substantial decay width we allow for
spectral distributions of the broadest vector mesons, the
$\rho_\mu$- and $K_\mu$-mesons. In channels involving the
$\rho_\mu-$ or $K_\mu-$meson the two-particle propagator is folded
with spectral functions obtained at the one-loop level describing
the decay processes $\rho_\mu \to \pi \,\pi$ and $K_\mu \to \pi\,
K$.

The Bethe-Salpeter interaction kernel $K_{\mu \nu}(\bar k,k;w)$ is
the sum of all two-particle irreducible diagrams. The scattering
problem decouples into thirteen orthogonal channels specified by
isospin ($I$), G-parity ($G$) and strangeness ($S$) quantum
numbers. From a field theoretic point of view once the interaction
kernel, $K_{\mu \nu}(\bar k,k;w)$,  and the two-particle
propagator, $G$, are specified the Bethe-Salpeter equation
determines the scattering amplitude $T_{\mu \nu}(\bar k, k;w)$.
However, in order to arrive at a scattering amplitude that does
not depend on the choice of interpolating fields at given order in
a truncation of the scattering kernel it is necessary to perform
an on-shell reduction.

Progress is made in \cite{LH02} by introducing an on-shell
equivalent effective interaction kernel $V$, together with three
off-shell interaction kernels $V_L,V_R$ and $V_{LR}$ where $V_R$
($V_L$) vanishes if the initial (final) particles are on-shell.
The interaction kernel $V_{LR}$ is defined to vanish if evaluated
with either initial or final particles on-shell. The latter
objects are defined by:
\begin{eqnarray}
K &=&V+ (1-V \cdot G)\cdot V_{L}+ V_{R}\cdot (1-G\cdot V)\,
\nonumber\\
&+& (1-V\cdot G)\cdot V_{LR}\cdot (1-G\cdot V) - V_R\cdot
\frac{1}{1-G\cdot V_{LR}}\cdot G \cdot V_L \,. \label{K-decomp}
\end{eqnarray}
The decomposition of the Bethe-Salpeter interaction kernel is
unique and can be applied to an arbitrary interaction kernel once
it is defined what is meant with the 'on-shell' part of any
two-particle amplitude. The latter we define as the part of the
amplitude that has a decomposition into the complete set of
projectors
\begin{eqnarray}
V_{\mu \nu}(\bar k,k;w) = \sum_{J, P,a,b}\,V^{(J
P)}_{ab}(\sqrt{s})\, {\mathcal Y}^{(J P)}_{\mu \nu,ab}(\bar q,q;w)
\,, \label{v-exp}
\end{eqnarray}
where the projectors carry good total angular momentum $J$ and
parity $P$. It is clear that performing a chiral expansion of $K$
and $V$ to some order $Q^n$ leads to a straight forward
identification of the off-shell kernels $V_L,V_R$ and $V_{LR}$ to
the same accuracy.

\subsection{Renormalization scheme and crossing symmetry}

Unlike in standard chiral perturbation theory the renormalization
of a unitarized chiral perturbation theory is non-trivial and
therefore requires particular care. The coefficient functions
$V^{(J P)}_{ab}(\sqrt{s})$ introduced in (\ref{v-exp}) are
evaluated in chiral perturbation theory and therefore standard
renormalization schemes are applicable. The on-shell part of the
scattering amplitude takes the simple form,
\begin{eqnarray}
&& T^{\rm on-shell}_{\mu \nu}(\bar k ,k ;w )  = \sum_{J,P}\,M^{(J
P)}(\sqrt{s}\,)\, {\mathcal Y}^{(J P)}_{\mu \nu}(\bar q, q;w) \,,
\nonumber\\
&& M^{(J P)}(\sqrt{s}\,) = \Big[ 1- V^{(J P)}(\sqrt{s}\,)\,J^{(J
P)}(\sqrt{s}\,)\Big]^{-1}\, V^{(J P)}(\sqrt{s}\,)\,, \label{}
\end{eqnarray}
with a set of divergent loop functions $J^{(J P)}(\sqrt{s}\,)$.
The crucial issue is how to renormalize the loop functions. In
\cite{LK02} it was suggested to introduce a physical scheme
defined by the renormalization condition,
\begin{eqnarray}
T_{\mu \nu}^{(J P)}(\bar k,k;w)\Big|_{\sqrt{s}= \mu } = V_{\mu
\nu}^{(J P)}(\bar k,k;w)\Big|_{\sqrt{s}= \mu } \,,
\label{ren-cond}
\end{eqnarray}
where the subtraction scale $\mu = \mu(I,S)$ depends weakly on
isospin and strangeness but is independent on $J^P$. It was argued
in \cite{LK02} that the optimal choice of the subtraction point
can be determined by the requirement that the scattering amplitude
is approximatively crossing symmetric. For the optimal subtraction
scales $\mu(I,S)$ we obtained,
\begin{eqnarray}
&& \mu(I,0) =  M_{\rho(770)}\,,\qquad \mu(I,\pm 1) =
M_{K(892)}\,,\qquad \mu(I,\pm 2) = M_{\rho(770)} \,. \label{}
\end{eqnarray}
Moreover it was demonstrated that the renormalization condition
(\ref{ren-cond}) is complete, i.e. the condition (\ref{ren-cond})
suffices to render the scattering amplitude finite. Before
discussing in some detail the choice of the subtraction points
$\mu (I,S)$ let us elaborate on the structure of the loop
functions. The merit of our scheme is that dimensional
regularization can be used to evaluate the latter ones. Here we
exploit the result that any given projector is a finite polynomial
in the available 4-momenta. This implies that the loop functions
can be expressed in terms of a log-divergent master function,
$I(\sqrt{s}\,)$, and reduced tadpole terms,
\begin{eqnarray}
&& J^{(J P)}(\sqrt{s}\,)= N^{(J
P)}(\sqrt{s}\,)\,\Big(I(\sqrt{s}\,) -I(\mu )\Big) \,,
\nonumber\\
&& I(\sqrt{s}\,)=\frac{1}{16\,\pi^2} \left(
\frac{p_{cm}}{\sqrt{s}}\, \left( \ln
\left(1-\frac{s-2\,p_{cm}\,\sqrt{s}}{m^2+M^2} \right) -\ln
\left(1-\frac{s+2\,p_{cm}\sqrt{s}}{m^2+M^2} \right)\right) \right.
\nonumber\\
&&\qquad \qquad + \left.
\left(\frac{1}{2}\,\frac{m^2+M^2}{m^2-M^2} -\frac{m^2-M^2}{2\,s}
\right) \,\ln \left( \frac{m^2}{M^2}\right) +1 \right)+I(0)\;,
\label{i-def}
\end{eqnarray}
where $\sqrt{s}= \sqrt{M^2+p_{cm}^2}+ \sqrt{m^2+p_{cm}^2}$. The
normalization factor $N_a^{(J P)}(\sqrt{s}\,)$ is a polynomial in
$\sqrt{s}$ and the mass parameters. In (\ref{i-def}) the
renormalization scale dependence of the scaler loop function
$I(\sqrt{s}\,)$ was traded in favor of a dependence on a
subtraction point $\mu$. The loop functions
$J^{(J,P)}(\sqrt{s}\,)$ are consistent with chiral counting rules
only if the subtraction scale $\mu \simeq M$ is chosen close to
the 'heavy' meson mass \cite{LK00,LK02}. Moreover it was shown
that keeping reduced tadpole terms in the loop functions leads to
a renormalization of s-channel exchange terms that is in conflict
with chiral counting rules. We emphasize that the projectors have
the important property that in the case of broad intermediate
states the implied loop functions follow from (\ref{i-def}) by a
simple folding with the spectral distributions of the two
intermediate states.

\begin{figure}[t]
\begin{center}
\includegraphics[width=12.0cm,clip=true]{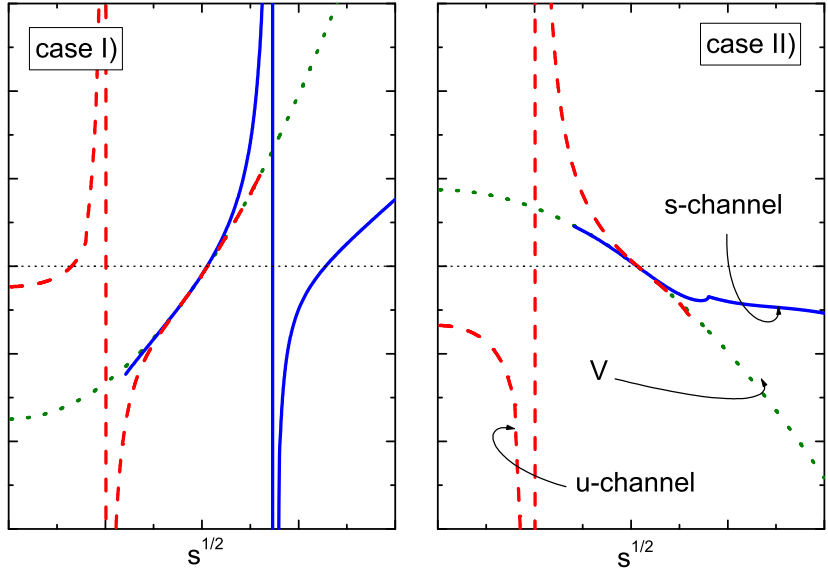}
\end{center}\vspace*{-1.3cm}
\caption{Typical cases of forward scattering amplitudes. The solid
(dashed) line
 shows the s-channel (u-channel)
unitarized scattering amplitude. The dotted lines represent the
amplitude evaluated at tree-level.} \label{fig:0}
\end{figure}

The renormalization condition (\ref{ren-cond}) reflects the basic
assumption our effective field theory is built on, namely, that at
subthreshold energies the scattering amplitudes may be evaluated
in standard chiral perturbation theory. Once the available energy
is sufficiently high to permit elastic two-body scattering a
further typical dimensionless parameter of order one that is
uniquely linked to the presence of a two-particle unitarity cut
arise. Thus it is sufficient to sum those contributions keeping
the perturbative expansion of all terms that do not develop a
two-particle unitarity cut. In order to recover the perturbative
nature of the subthreshold scattering amplitude the subtraction
scale $ M-m < \mu < M+m$ must be chosen in between the s- and
u-channel elastic unitarity branch points \cite{LK02}.  It was
suggested that s-channel and u-channel unitarized amplitudes
should be glued together at subthreshold kinematics \cite{LK02}. A
smooth result is guaranteed if the full amplitudes match the
interaction kernel $V$ close to the subtraction scale $\mu$ as
imposed by (\ref{ren-cond}). In this case the crossing symmetry of
the interaction kernel, which follows directly from its
perturbative evaluation, carries over to an approximate crossing
symmetry of the full scattering amplitude. This construction
reflects our basic assumption that diagrams showing an s-channel
or u-channel unitarity cut need to be summed to all orders
typically at energies where the diagrams develop their imaginary
part. In Fig. \ref{fig:0} we demonstrate the quality of the
proposed matching procedure as applied for typical forward
scattering amplitudes. The figure clearly illustrates the smooth
matching of s-channel and u-channel iterated amplitudes at
subthreshold energies.

Given the subtraction scales as derived above the leading-order
calculation is parameter-free. Of course chiral correction terms
lead to further so far unknown parameters  which need to be
adjusted to data. Within the $\chi-$BS(3) approach such correction
terms enter the effective interaction kernel $V$ rather than
leading to a change of the subtraction scales.

\section{Results}

We present out results on s-wave scattering of Goldstone bosons
off vector mesons using the leading order chiral SU(3) Lagrangian.
Meson resonances with quantum number (I,S) and $J^P\!=\!1^+$
manifest themselves as poles in the corresponding scattering
amplitudes $M_{ab}^{JP}(\sqrt{s})$.

\begin{figure}[t]
\begin{center}
\includegraphics[width=14.5cm,clip=true]{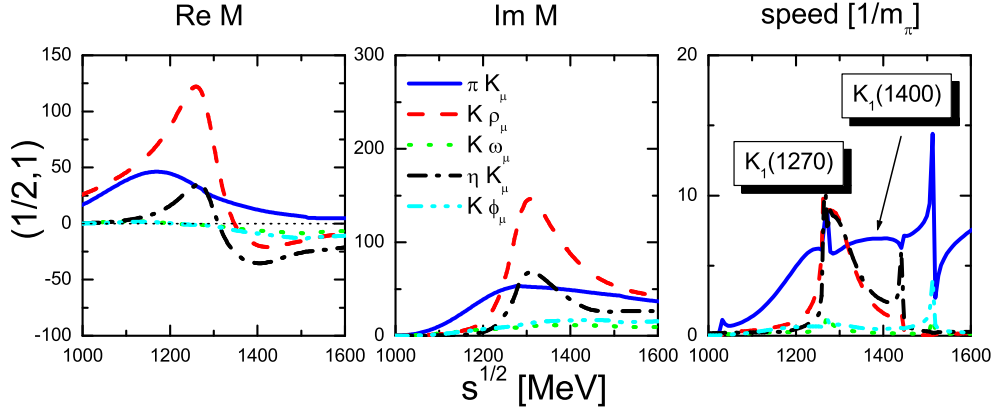}
\end{center}\vspace*{-1.3cm}
\caption{Scattering amplitudes and speeds for meson resonances
with $J^P\!=\!1^+$ and $(I,S)=(\frac{1}{2},1)$. } \label{fig:1}
\end{figure}

\begin{figure}[t]
\begin{center}
\includegraphics[width=14.5cm,clip=true]{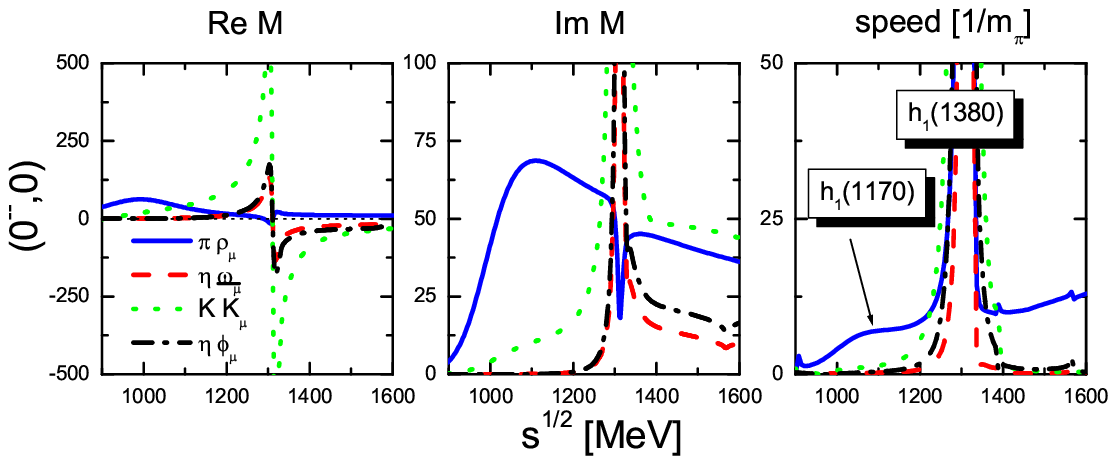}
\end{center}\vspace*{-1.3cm}
\caption{Scattering amplitudes  and speeds for meson resonances
with $J^P\!=\!1^+$ and $(I^G,S)=(0^-,0)$. } \label{fig:2}
\end{figure}

\begin{figure}[t]
\begin{center}
\includegraphics[width=14.5cm,clip=true]{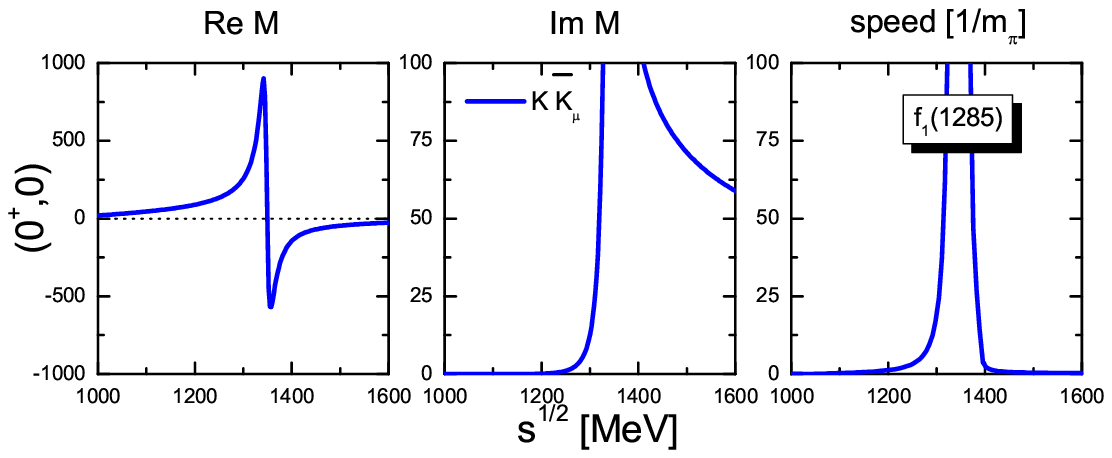}
\end{center}\vspace*{-1.3cm}
\caption{Scattering amplitudes  and speeds for meson resonances
with $J^P\!=\!1^+$ and $(I^G,S)=(0^+,0)$.} \label{fig:3}
\end{figure}

\begin{figure}[t]
\begin{center}
\includegraphics[width=14.5cm,clip=true]{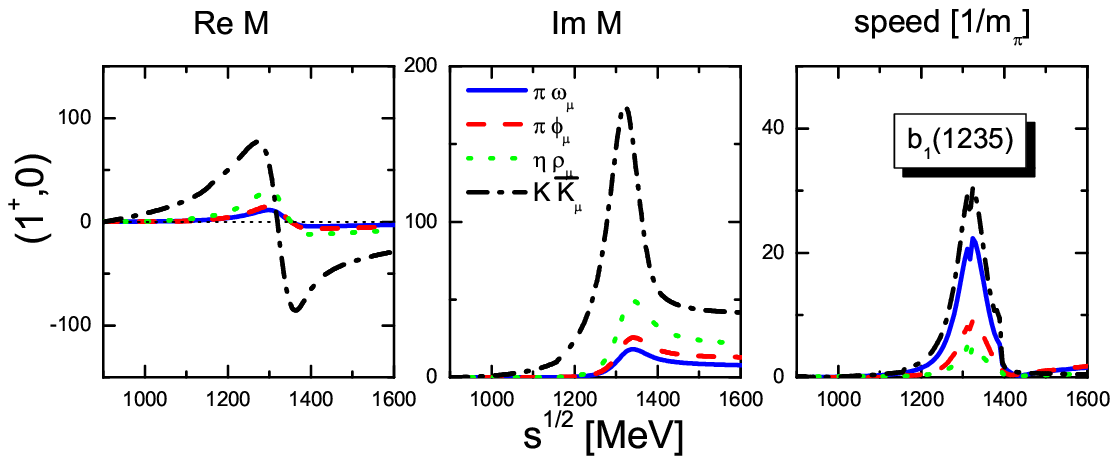}
\end{center}\vspace*{-1.3cm}
\caption{Scattering amplitudes  and speeds for meson resonances
with $J^P\!=\!1^+$ and $(I^G,S)=(1^+,0)$. } \label{fig:4}
\end{figure}

\begin{figure}[t]
\begin{center}
\includegraphics[width=14.5cm,clip=true]{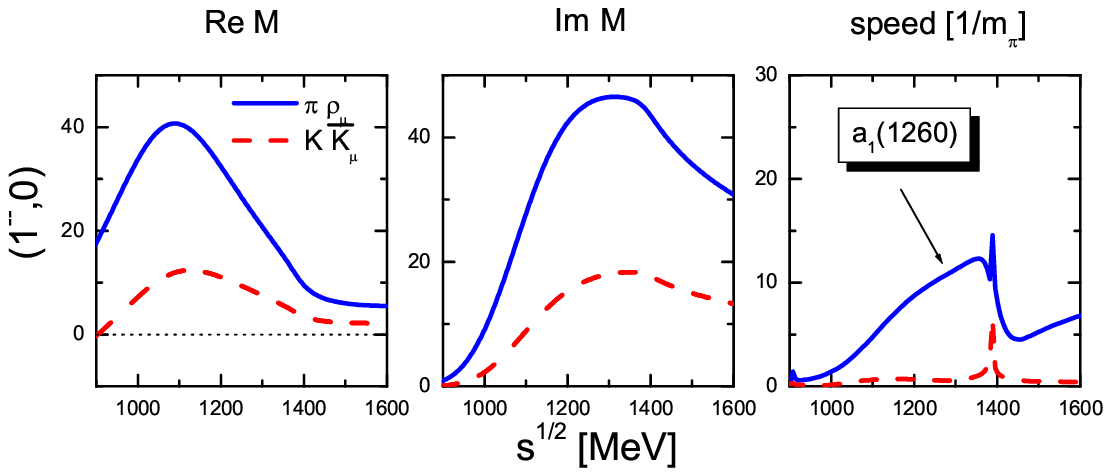}
\end{center}\vspace*{-1.3cm}
\caption{Scattering amplitudes  and speeds for meson resonances
with $J^P\!=\!1^+$ and $(I^G,S)=(1^-,0)$. } \label{fig:5}
\end{figure}

To study the formation of meson resonances we generate speed plots
(see \cite{LK03}). The merit of producing speed plot lies in a
convenient property of the latter allowing a straight forward
extraction of resonance parameters. Assume that a coupled-channel
amplitude $M_{ab}(\sqrt{s}\,)$ develops a pole of mass $m_R$, with
\begin{eqnarray}
&& M_{ab}(\sqrt{s}\,) = -\frac{
g^*_a\,g^{\phantom{*}}_b\,m_R}{\sqrt{s}-m_R + i\,\Gamma/2}
\,,\qquad \Gamma_a =
\frac{|g_a|^2}{4\,\pi}\,|p^{(a)}_{cm}|\,N_a(m_R)\,,
\label{def-res}
\end{eqnarray}
where the total resonance width, $\Gamma$, is given by the sum of
all partial widths. The normalization factor $N (M_R)$ in
(\ref{def-res}) is identical to the one entering the form of the
loop functions in (\ref{i-def}). The speed plots take a maximum at
the resonance mass $\sqrt{s}=m_R$, with
\begin{eqnarray}
&& {\rm Speed}_{aa}(m_R\,) = \Bigg\{
\begin{array}{ll}
2\,\frac{\Gamma_a}{\Gamma^2}\,\Big|2\,\sum_{c}\,\frac{\Gamma_c}{\Gamma}
-1\Big| \,
\qquad \qquad  \,{\rm if} \; a= {\rm open}\\
2\,\frac{\Gamma_a}{\Gamma^2}\,\Big|2\,
\sum_{c}\,\frac{\Gamma_c}{\Gamma} -i\,\Big| \, \qquad \qquad
\,{\rm if} \; a= {\rm closed} \,,
\end{array}
\nonumber\\
&& \Gamma = \sum_{a={\rm open}} \Gamma_a \,. \label{speed-an}
\end{eqnarray}
The result (\ref{speed-an}) clearly demonstrates that the speed of
a resonance in a given open channel $a$ is not only a function of
the total width parameter $\Gamma$ and the partial width
$\Gamma_a$. It does depend also on how strongly closed channels
couple to that resonance. This is in contrast to the delay time
of a resonance for which closed channels do not
contribute.

To explore the multiplet structure of the resonance states we
study first the 'heavy' SU(3) limit \cite{Granada,Copenhagen} with
$m_{\pi, K, \eta} \simeq 500$ MeV and $m_{\rho, \omega, K^*,
\phi}\simeq 900$ MeV. In this case all resonance states turn into
bound states forming two degenerate octets and one singlet of the
SU(3) flavor group with masses  1367 MeV and 1289 MeV
respectively. The latter numbers are quite insensitive to the
precise value of the subtraction scale. For instance increasing
(decreasing) the subtraction scale by 20 $\%$ away from its
natural value the octet bound-state mass comes at 1383 MeV (1353
MeV). Our result is a direct reflection of the Weinberg-Tomozawa
interaction,
\begin{eqnarray}
8 \otimes 8 = 1 \oplus 8 \oplus 8 \oplus 10 \oplus
\overline{10}\oplus 27 \,, \label{}
\end{eqnarray}
which predicts attraction in the two octet and the singlet
channel. This finding is analogous to the results of
\cite{Granada} that found two degenerate octet and one singlet
state in the SU(3) limit of meson-baryon scattering with
$J^P\!=\!\frac{1}{2}^-$. Taking the 'light' SU(3) limit
\cite{Copenhagen} with $m_{\pi, K, \eta} \simeq 140$ MeV and
$m_{\rho, \omega, K^*, \phi}\simeq 700$ MeV we do not observe any
bound-state nor resonance signals anymore. A further interesting
limit to study is $N_c \to \infty$. Since the scattering kernel is
proportional to the $f^{-2}\sim N_c^{-1}$ the interaction strength
vanishes in that limit and no resonances are generated.

Figs. \ref{fig:1}-\ref{fig:5} show the resonance spectrum that
arises using physical masses using realistic spectral
distributions for the broad vector mesons. The figures show real
and imaginary parts of the scattering amplitudes as well as the
associated speed plots. Clear signals in the speed plots of the
$({\textstyle{1\over 2}},\pm 1)$, $(0^\pm,0)$ and $(1^\pm,0)$
channels are seen. No resonance is found in the remaining
channels. The resonances can be unambiguously identified with the
axial-vector meson resonances ($h_1(1170), h_1(1380), f_1(1285),
a_1(1260), b_1(1235), K_1(1270), K_1(1400)$).

In the 'heavy' SU(3) limit the $({\textstyle{1\over 2}},\pm 1)$
channel shows two bound states reflecting the presence of two
degenerate octet states. Using physical masses the degeneracy is
lifted and a narrow state at 1263 MeV and a broad state at about
1300 MeV arise. The effect of using realistic spectral
distributions for the intermediate $\rho_\mu$- and $K_\mu$-mesons
is demonstrated in Fig. \ref{fig:1}. The resonance signal in the
speed plots becomes much clearer since using spectral
distributions for the broad intermediate states smears away the
square-root singularities in the speeds at the corresponding
thresholds. Our result is quite consistent with the empirical
properties of the $K_1(1270)$ meson. It has a width of about 90
MeV and decays dominantly into the $K\,\rho_\mu-$channel
\cite{PDG02}. The second much broader state is assigned a width of
about 175 MeV resulting almost exclusively from its decay into the
$\pi \,K_\mu-$channel \cite{PDG02}.

Similarly, in the heavy SU(3) limit the $(0^-,0)$ channel shows
two bound states associated with a singlet and an octet state.
Using physical masses a broad state at about 1100 MeV and a narrow
state at 1303 MeV should be identified with the $h_1(1170)$ and
$h_1(1380)$ resonance. Here we assign the $h_1(1380)$-resonance,
for which its quantum numbers except its parity and angular
momentum $J^P=1^+$ are unknown, the isospin and G-parity quantum
numbers $I^G=0^-$. This is a clear prediction of the chiral
coupled-channel dynamics. The latter state has so far been seen
only through its decay into the $K\,\bar K_\mu$- and $\bar
K\,K_\mu -$channels \cite{PDG00}. Its small width of about 80 MeV
\cite{PDG00} is consistent with the narrow structure seen in Fig.
\ref{fig:2}. The second resonances state in Fig. \ref{fig:2} is
most clearly seen in the $\pi \rho_\mu-$channel. This is
consistent with the empirical properties of the $h_1(1170)$
resonance which so far has been seen only through its $\pi
\rho_\mu-$decay leading to a large width of about 360 MeV
\cite{PDG02}.

The $(0^+,0)$-speed (see Fig. \ref{fig:3}) shows a bound-state at
mass 1341 MeV a value somewhat above the mass of the $f_1(1285)$
resonance. Using a spectral distribution for the $K_\mu(892)$ in
the intermediate states $K\,\bar K_\mu$ and $\bar K\,K_\mu $
states a narrow resonance appears. Its width of about 10 MeV is a
factor two smaller than the empirical value \cite{PDG02}. The
$(1^+,0)$-speed of Fig. \ref{fig:4} shows a resonance at 1310 MeV
to be identified with the $b_1(1235)$ resonance. From the maximum
of the imaginary part of the scattering amplitudes at the
resonance peak one can directly read off ratios of coupling
constants. Fig. \ref{fig:4} clearly demonstrates that the smallest
coupling constant is predicted for the $\pi\, \omega_\mu
-$channel. Nevertheless, the hadronic decay of the $b_1(1235)$ is
completely dominated by the $\pi\,\omega_\mu-$channel. This is a
simple consequence of phase-space kinematics. The widths of the
resonance as indicated by Fig. \ref{fig:4} is quite compatible
with the empirical value of about 140 MeV \cite{PDG02}.  The
$a_1(1260)$ resonance is found in the $(1^-,0)$-speed of Fig.
\ref{fig:2} as a broad peak with a mass of about 1300 MeV.
Empirically its width is estimated to be about 250-600 MeV
\cite{PDG02} resulting from its decay into the
$\pi\,\rho-$channel.

The structure of the $h_1(1380),f_1(1285)$ and $b_1(1235)$
resonances as predicted by chiral coupled-channel dynamics is
quite intriguing since those resonances couple dominantly to the
$K\,\bar K_\mu-$channel. This implies that the latter channel is
the driving force that generates these resonances dynamically.
This finding is very much analogous to the structure of the scalar
$f_0(980)$ resonance that strongly couples to the $K \bar
K-$channel and emphasizes the importance of the chiral SU(3)
symmetry even for non-strange resonances. It should be emphasized
that the results obtained here at leading order can be improved
further by incorporating chiral correction terms into the
analysis. In view of the remarkable success of the leading order
Weinberg-Tomozawa interaction one would expect a rapidly
converging expansion.

\end{document}